# Experimental demonstration of directive Si$_3$N$_4$ optical leaky wave antenna with semiconductor perturbations at near infrared frequencies

Qiancheng Zhao[1], Caner Guclu[1], Yuewang Huang[1], Salvatore Campione [1,#], Filippo Capolino[1], Ozdal Boyraz[1]

[1]Department of Electrical Engineering and Computer Science, University of California-Irvine, Irvine, CA, USA, 92697

[#]Current address: Center for Integrated Nanotechnologies, Sandia National Laboratories, Albuquerque, NM, USA, 87185

**ABSTRACT**

Directive optical leaky wave antennas (OLWAs) with tunable radiation pattern are promising integrated optical modulation and scanning devices. OLWAs fabricated using CMOS-compatible semiconductor planar waveguide technology have the potential of providing high directivity with electrical tunability for modulation and switching capabilities. We experimentally demonstrate directive radiation from a silicon nitride (Si$_3$N$_4$) waveguide-based OLWA. The OLWA design comprises 50 crystalline Si perturbations buried inside the waveguide, with a period of 1 μm, each with a length of 260 nm and a height of 150 nm, leading to a directive radiation pattern at telecom wavelengths. The measured far-field radiation pattern at the wavelength of 1540 nm is very directive, with the maximum intensity at the angle of 84.4° relative to the waveguide axis and a half-power beam width around 6.2°, which is consistent with our theoretical predictions. The use of semiconductor perturbations facilitates electronic radiation control thanks to the refractive index variation induced by a carrier density change in the perturbations. To assess the electrical modulation capability, we study carrier injection and depletion in Si perturbations, and investigate the Franz-Keldysh effect in germanium as an alternative way. We theoretically show that the silicon wire modulator has a -3 dB modulation bandwidth of 75 GHz with refractive index change of 3×10$^{-4}$ in depletion mode, and 350 MHz bandwidth with refractive index change of 1.5×10$^{-2}$ in injection mode. The Franz-Keldysh effect has the potential to generate very fast modulation in radiation control at telecom wavelengths.

## 1. INTRODUCTION

Optical antennas direct light, focus energy, and enhance light-matter interaction. Therefore they are promising for applications such as solar cells, planar imaging [1], space-division multiplexing [2], [3] devices, and tunable grating couplers [4]. An optical leaky wave antenna (OLWA) is a device that radiates a light wave into the surrounding space from an integrated dielectric waveguide by means of a guided leaky wave mode [5]. The importance of leaky-waves in achieving directive radiation and enhanced transmission in corrugated thin plasmonic films was reported in [6]. Recent studies explore operational regimes from dielectric-based OLWA to metal-based leaky plasmon waves [7] such as subwavelength aperture in silver films [8], and nanoparticle structures [9], [10].

The OLWA presented in this paper consists of a dielectric waveguide with buried silicon perturbations that hosts a slowly attenuating leaky-wave mode as reported extensively in [11]. The periodicity of the silicon perturbations of the waveguide introduces spatial Floquet harmonics of the fundamental guided mode, one of which (usually the -1 harmonic) is engineered as a fast wave that leads to leaky-wave radiation. A very slowly attenuating leaky wave is realized by designing silicon inclusions with a small filling fraction, thus a very directive radiated beam is achieved. We have chosen a silicon platform to realize the leaky-wave antenna device. The versatile silicon platform is convenient not only in communications, but also in sensing [12], imaging [13], and microfluidic devices [14]. Especially the silicon-on-insulator (SOI) platform, which reached a mature stage for integrated silicon photonics, has been widely used to deliver chip-scale passive photonic devices such as switches, modulators, detectors, and nonlinear devices [15] with CMOS fabrication compatibility [16]. In particular the electronic tunability of the optical parameters of silicon via plasma dispersion effect [17], [18] or Franz-Keldysh effect [19] renders SOI the ideal platform for optical antennas that can facilitate electronic beam control.

In this paper, we present the experimental demonstration of a dielectric CMOS-compatible OLWA. Readers may refer to previous works [5], [11], [20] for further information on the theoretical design of OLWAs. Section 2 presents the fabrication of the OLWA. Section 3 describes the radiation performance of a fabricated OLWA, followed by the theoretical study on electrical modulation capability in section 4.

## 2. FABRICATION OF THE OLWA

UniBond SOI wafer based on the Smart-Cut process is chosen as the substrate platform due to its advantages of high crystalline quality and availability of thick silicon dioxide layer [21]. The silicon dioxide layer on which the waveguide is deposited is 1 µm thick so as to prevent mode leaking into the silicon substrate. The processing of silicon perturbation starts with thinning the silicon top layer to 150 nm using lower power boron chloride, chlorine and oxygen plasma etching. An ASML 5500 optical stepper is used to pattern silicon perturbations. The stepper reticle scaled up by 5 times is first fabricated by Compugraphics. A backside anti-reflection coating layer (BARC) is then coated under the photoresist layer to eliminate interference from reflection and hence to achieve better resolution. The silicon layer is etched using Fluorine RIE plasma. After etching, the photoresist and backside anti-reflection coating layer are removed by high power oxygen plasma.

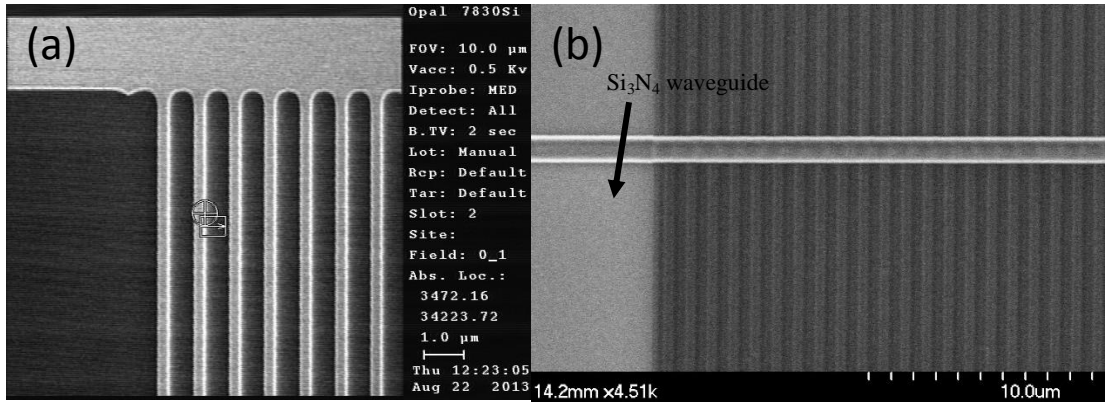

Figure 1. (a) Scanning electron microscopy (SEM) image of the fabricated silicon perturbations. (b) SEM picture of the fabricated silicon nitride waveguide sitting on the periodic silicon perturbations.

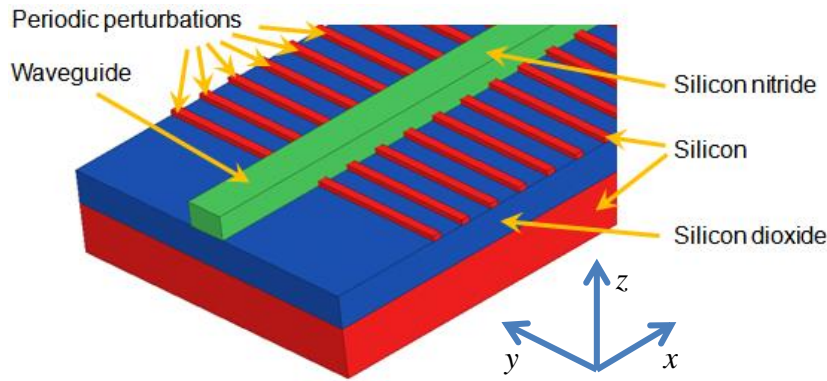

Figure 2. Three dimensional illustration of the OLWA made of a silicon nitride waveguide over periodic silicon perturbations for possible electronic control.

The fabricated silicon wire perturbations are shown in Fig. 1a. The narrow vertical lines are silicon wire perturbations which are attached to a silicon "skeleton" to prevent the perturbation wires to be peeled off. The width of each silicon wire is measured to be in average 260 nm. The spatial periodicity is ~1 µm. The fabricated silicon nitride waveguide shown in Fig. 1b, extending horizontally in the SEM picture, is deposited on top of the silicon wires patterned in the previous step. Low-pressure chemical vapor deposition (LPCVD) is utilized with a stoichiometric recipe. After deposition of the silicon nitride film, the SPR-700 is coated to form a photoresist film of 1 μm thickness. The photoresist strips serve as a mask for silicon nitride etching, are fabricated by standard lithography process. Then the silicon nitride waveguide is etched by using Oxford Fluorine RIE etcher. The fabricated silicon nitride waveguide is measured to have 1 μm width and 735 nm height.

## 3. ANTENNA CHARACTERIZATION

### 3.1 Experimental setup

The OLWA is designed with the capability of highly directional radiation at telecom wavelength both in upward and downward directions off the wafer plane. According to the full-wave simulations, downward radiation has higher intensity, and yet is difficult to detect, because the bottom silicon surface is rough and the substrate layer is thick. Therefore only the top radiation is measured here. Since this is a waveguide-based device, butt-coupling is employed to launch light in the waveguide. A tapered lens fiber is employed to focus light into the waveguide facet to achieve high coupling efficiency. Due to detection difficulty at near-infrared at low power, a red laser is used as a visible indicator for coupling. When the tapered lens fiber is at the right position, a light spot can be observed at the silicon corrugations in the waveguide, indicating leakage wave.

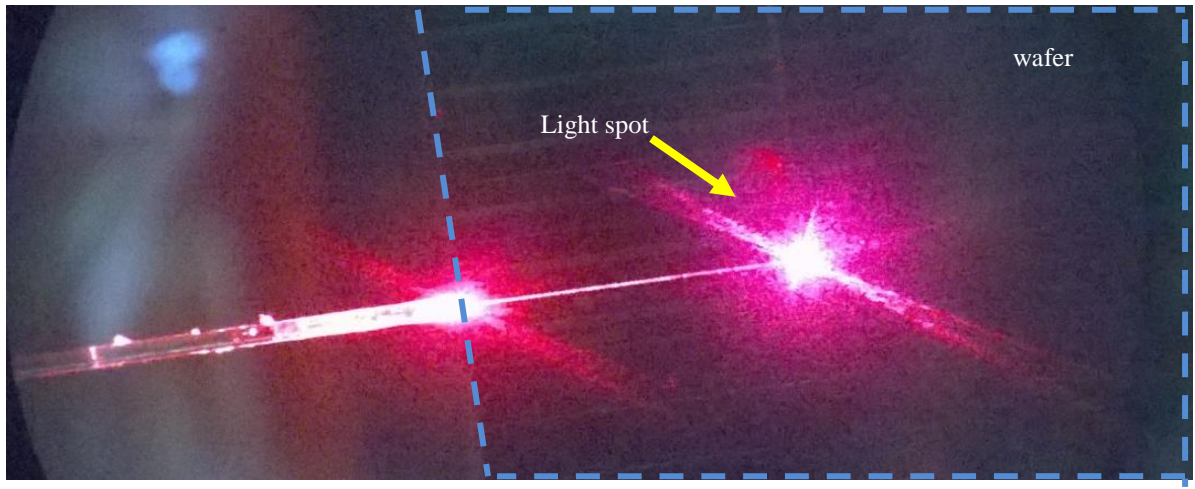

Figure 3. Microscope image of red light emission at the silicon corrugation position in the waveguide. The light is butt-coupled by a tapered lens fiber, and propagates in the silicon nitride waveguide. The light spot indicates wave leakage. The dashed line circles the region of the wafer, located at the right part of the figure.

The OLWA excitation is then repeated at 1540 nm. Since the device radiates a narrow beam into space with weak power, a fiber tip is employed to detect the radiated power above the OLWA. Due to the fiber's small numerical aperture around 0.12, corresponding to a collection angle of ±7°, the fiber is a narrow beam detector. Maximum detection efficiency occurs only when the fiber is pointing to the emitting point of the antenna. Thus an angle controller is used to adjust the fiber angle. An optical power meter is used to monitor the power coupled by the fiber tip. The schematic of the experiment setup is illustrated Fig. 4.

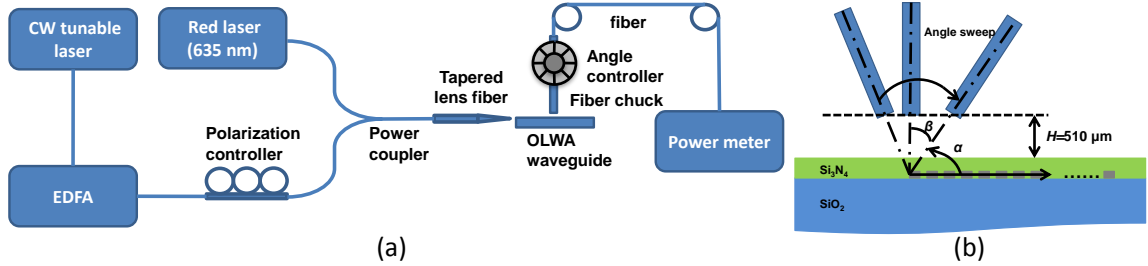

Figure 4. (a) Schematic of OLWA radiation testing experiment setup. (b) Schematic of the OLWA angular sweep test experiment. The red thick arrow indicates the waveguide propagation direction. The emission point is estimated at the very first silicon perturbations location. The angle between the radiation observation and waveguide propagation direction is denoted as α. The angle between the probing fiber axis and waveguide surface normal is denoted as β. Various β are tested in the experiment.

### 3.2 Experiment methods

The radiation pattern is characterized versus an angle α, defined as the angle with respect to the guided wave propagation direction as shown in Fig. 4b. To perform angular measurements, the fiber tip is suspended on top of the waveguide. The vertical distance between fiber tip and the waveguide surface is estimated to be approximately $H$ = 510 μm. During the angle sweeping, the measuring fiber tip is rotated by the angle rotator with step size Δβ= 1°. As the angle β is varied, the probe is maintained at the vertical distance $H$ = 510 μm, and the horizontal position is compensated by stage adjustment via fine tuning. The horizontal position is adjusted to achieve maximum power reception which takes place when the fiber tip points to the OLWA. The radiation pattern is calculated by collecting the power value at each fiber tip angle, along with the horizontal distance between the OLWA and the tip. The data is then post-processed for compensation of different free-space path losses at different measurement angles. Moreover, by assuming a linear power drift by time, the recorded power values are adjusted to compensate for the impact of power drift.

### 3.3 Experiment results

The radiation pattern versus angle α, measured from the propagation direction of the injected wave, is plotted in Fig. 5. The curve has a single dominant peak in the angle range between 65° to 112°, and all the side lobes are -5 dB below the main peak, indicating a directive emission. The low contrast between the main lobe and the side lobe may result from noise and limited input power. A higher input power can distinguish the main lobe more effectively. The main lobe radiation angle $α_{max}$ is found by averaging $α_{max} = (α_{-3dB\_left} + α_{-3dB\_right})/2$ to reduce noise effect, where $α_{-3dB}$ denotes the angle at which radiation intensity drops to one half of the maximum one. The measured far-field radiation pattern at the wavelength of 1540 nm has its maximum intensity at the angle of 84.4° and a half-power beam width around 6.2°, which is consistent with our theoretical predictions [5].

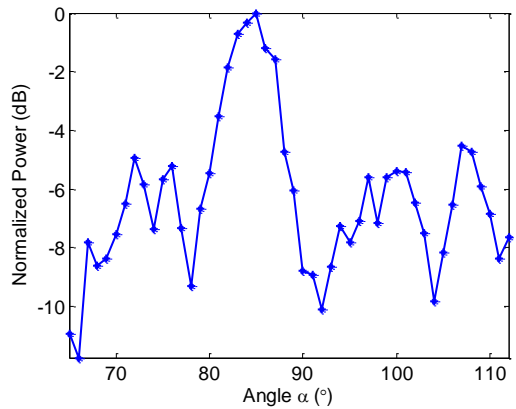

Figure 5. Experimental result: Measured normalized radiation pattern (in the *x-z* plane, see Fig. 6) versus angle α at wavelength 1540 nm. Note the radiation peak at 84.4 degrees.

## 4. ELECTRIC MODULATION

The far-field radiation pattern can be tuned by changing the refractive index of silicon corrugations via plasma dispersion effect [22]. The refractive index variation induced by carrier density can be achieved by either carrier injection or carrier depletion. In this part, we explore the concept of electrical modulation of leaky wave radiation by varying the carrier concentration inside the silicon perturbations.

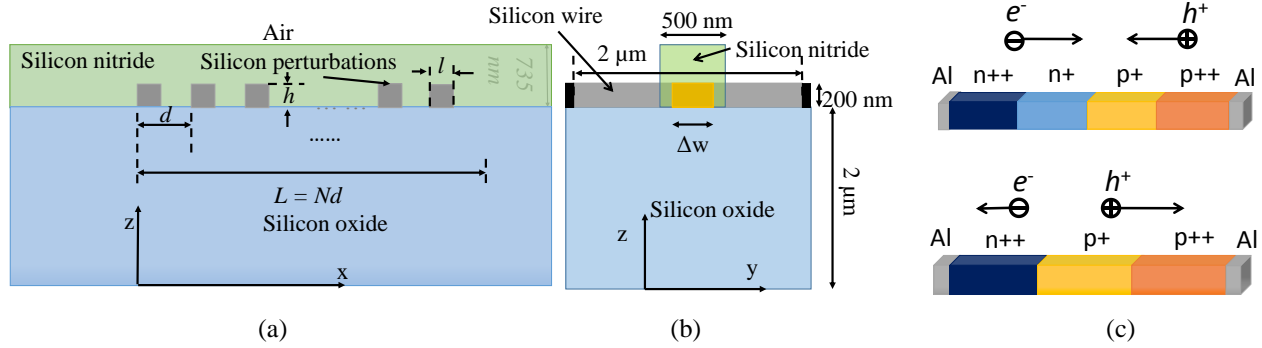

Figure 6. (a) Lateral view of the OLWA geometry made of a silicon nitride waveguide along *x*. The radiating part of the OLWA has length *L* (b) Front view of the OLWA, showing the silicon nitride waveguide with a cross-section of 0.5 μm × 0.5 μm and one silicon wire. Silicon wires, 2 μm long with a square cross section 200 nm × 200 nm, are embedded in the silicon nitride waveguide, sitting on a 2 μm thick SOI. The yellow region is the modulation region, with length in the *y* direction equal to Δ*w*. The aluminum contacts (black) are connected to the two ends. (c) Structure of the *p-n* junction in carrier injection (upper) and *p-i-n* junction in carrier depletion (lower). The intrinsic region in *p-i-n* junction is actually *p+* region. Compared the heavily doped *n++* and *p++*, it is "intrinsic".

In general, the larger the silicon perturbations are, the stronger the modulation effect is. However, this is only an approximation, because larger silicon perturbations may cause stronger leakage and hence wider radiating beams. We investigate here an example consisting of an OLWA with an increased filling fraction of Si domain in a unit cell, as shown in Fig. 6a. The silicon nitride waveguide is 500 nm high by 500 nm wide. The silicon perturbation wires are 0.2 μm × 2 μm × 0.2 μm. Each perturbation wire can be viewed as an individual modulator. Since the wires are periodic in space, the analysis applied to one wire modulator is also suitable for all others. Fig. 6b and Fig. 6c illustrate the geometry of perturbations that can accommodate electrical controls. The upper structure in Fig. 6c is for carrier injection, whereas the lower one is for carrier depletion.

The geometry used in injection mode is a *p-n* junction structure with symmetric doping profile of $N_a = N_d = 10^{16}$ cm$^{-3}$, and with a length of 0.5 μm in both *n+* and *p+* regions. The doping concentrations and lengths are chosen to make the space charge region as wide as the nitride waveguide width. The two ends of the silicon wire are heavily doped to $10^{19}$ cm$^{-3}$ to form good ohmic contact with the aluminum nodes.

Similarly, the optimization of the electrical property in depletion mode gives a structure like a *p-i-n* diode. P-type silicon is preferred as the "intrinsic" part due to higher impact on refractive index. In this way, doping concentration is balanced between refractive index change, modulation width and breakdown voltage. A moderate doping level of $10^{17}$ cm$^{-3}$ is found to be optimum. The boundary of the depletion region is defined where the carrier density drops to 1/10 of the doping concentration, and the modulation width is defined as the difference between maximum and minimum depletion widths. To make the modulation region symmetric to the waveguide, the *n++* region has a length of 0.8 μm with $10^{19}$ cm$^{-3}$ dopant concentration. The *p++* region, 0.5 μm long, is also highly doped to $10^{19}$ cm$^{-3}$. The silicon-silicon nitride surface recombination velocity is chosen to be $S_{eff} = 10^2$ cm/s for stoichiometric silicon nitride [23]. The theoretical refractive index change is 3.4×10$^{-4}$ with a modulation width of 0.178 μm in DC bias condition.

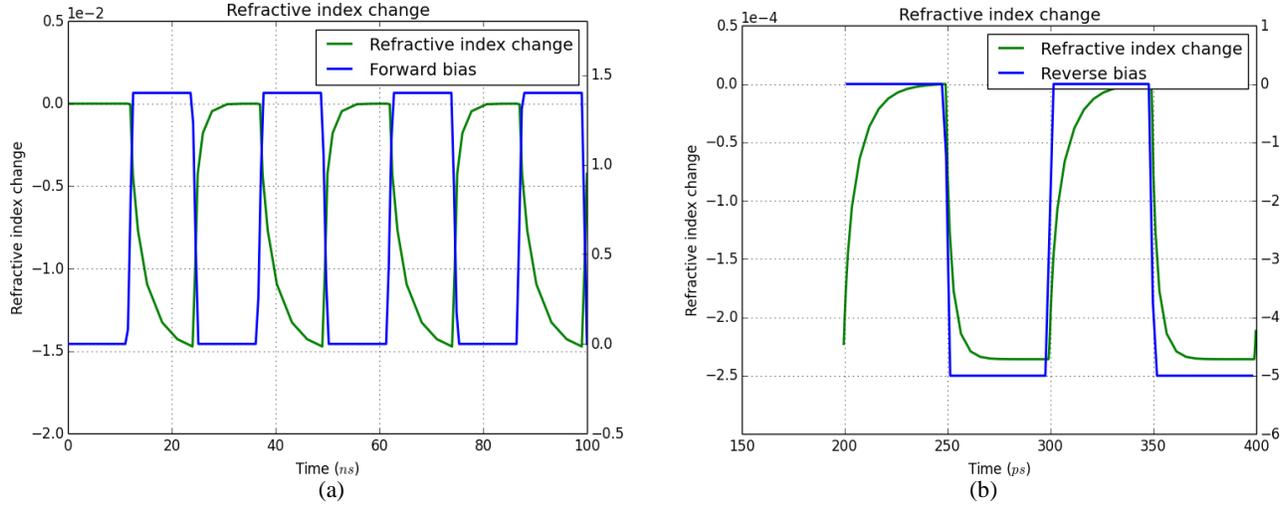

Figure 7. (a) Simulation of the average refractive index change in *injection mode* using a 40 MHz square wave with $V_{pp}$ of 1.4 V and offset 0.7 V. (b) Simulation of the average refractive index change in *depletion mode* using a 10 GHz square wave with $V_{pp}$ of 5 V and offset -2.5 V. The fluctuation of the refractive index change becomes stable after some time, thus the initial simulation before 200 ps are not included.

Transient simulations show a refractive index step response in both Fig. 7a and Fig. 7b, for the *injection* and *depletion* modes of operation, respectively. The average refractive index change is simulated by using a 40 MHz square wave for carrier injection mode (Fig. 7a) and a 10 GHz square wave for carrier depletion mode (Fig. 7b). The 10% - 90% rise time and fall time for injection mode are 12 ns and 3 ns, respectively. The rise and fall time for depletion mode are about 20 ps and 10 ps, respectively. It is apparent that the depletion mode operation can be three orders of magnitude faster than the injection mode which suffers from minority carrier lifetime issues. The discrepancy between rise time and fall time is attributed to the electric field dependent carrier velocity, and can be overcome by a pre-emphasized signal [24]. Although carrier injection mode has a drawback of low speed, it outweighs depletion mode on refractive index change. The *Δn* in injection mode can be as high as $1.5 \times 10^{-2}$, while that in depletion mode is only $2.5 \times 10^{-4}$.

The modulation width for AC response in *carrier depletion* mode is investigated to explore a possible device bandwidth. High frequency modulation widths are normalized to DC width for a fair comparison as illustrated in Fig. 8. It can be concluded that higher doping concentration leads to a higher bandwidth of operation. For $N_a = 1 \times 10^{17}$ cm$^{-3}$ in the "intrinsic" region (p+), the device's bandwidth is 75 GHz. To compare with small signal bandwidth which is estimated by -3 dB cutoff frequency, the junction capacitance is extracted from simulation. The capacitance is $4.3 \times 10^{-17}$ F, and the resistance is approximately $3.86 \times 10^4$ Ω. Therefore the *RC* cutoff frequency is 96 GHz, which is larger than the modulation width bandwidth. This may be attributed to the carrier velocity saturation, carrier generation and recombination, and other factors that are not included in the *RC* small signal model which can only give an upper limit of bandwidth. In the *carrier injection* mode, the junction capacitance is about 0.05 pF, while the silicon wire resistance is estimated to be 9 KΩ when it is half injected. The corresponding cutoff frequency is about 350 MHz. However, this bandwidth can be further optimized to operate at higher speeds.

*Carrier depletion* mode is limited by refractive index variation, so Franz-Keldysh effect on germanium is to be explored as an alternative approach. Since Franz-Keldysh effect is weak in pure silicon at telecommunication wavelength [22], germanium is explored as an alternative perturbation material. By varying the electric field in a germanium wire diode, absorption coefficient is affected, and the refractive index is altered according to Kramers-Kronig relations. The refractive index change can be as large as 0.01 near 1550 nm wavelength, when the applied electric field is 70 kV/cm. Since the Franz-Keldysh effect does not involve minority carriers, the modulation bandwidth is estimated to exceed far beyond *p-i-n* diode configuration.

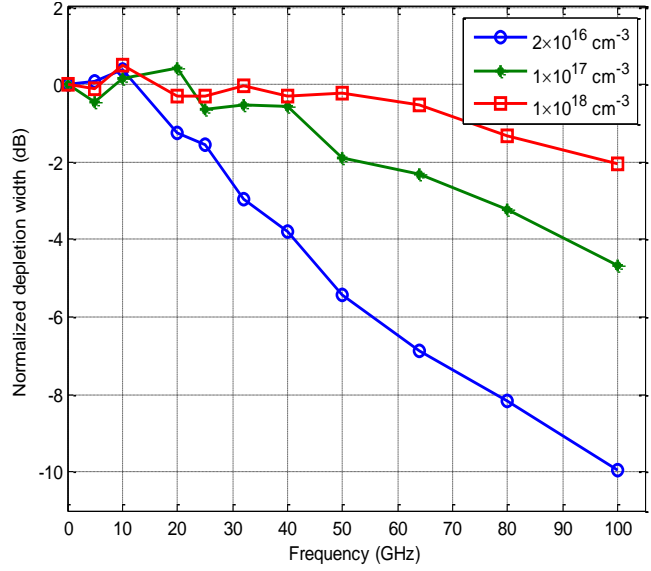

Figure 8. AC response of normalized modulation width with 5 Vpp square wave input. Three devices with different doping concentrations, $N_a = 2\times10^{16}$ cm$^{-3}$, $1\times10^{17}$ cm$^{-3}$, and $1\times10^{18}$ cm$^{-3}$, in the "intrinsic" region are compared. Depletion widths are normalized to DC bias conditions.

Based on the carrier dynamics of the three dimensional (3-D) geometry of the silicon perturbations (Figs. 1, 2 and 6), we assess the impact of carrier density modulation on the far-field radiation pattern. For simplicity, we investigate a 2-D OLWA model as was done in initial part of [5], using a commercial finite-element-method solver (COMSOL), where the Si domain is modeled with the carrier density-dependent refractive index derived from the 3-D modulator model above. We adopt the silicon wires geometry (200 nm × 200 nm in *x-z* plane), refractive index, carrier density change and refractive index change from the modulation model. The OLWA considered consists of a silicon nitride waveguide with a height of 500 nm and a length of 150.4 μm. We choose a thinner waveguide to increase of volume ratio of silicon perturbations to waveguide, so that the modulation will be more effective to change the effective refractive index in the perturbation region. There are 150 silicon perturbations with 200 nm × 200 nm cross section in the *x-z* plane, and *d* =920 nm spatial periodicity along the *x* direction (Fig. 6). The simulation result in Fig. 9 shows that the far field radiation pattern in the *x-z* plane, at 1550 nm has one single sharp peak at 87.7°, and a directivity of 54.65 dB with a beam width of 0.5°. This is significantly narrower than the 3-D OLWA case. It is an expected feature due to the weaker scattering by the perturbations in the simplified 2-D OLWA model than the perturbation scattering in the 3-D case. Therefore, the 2-D model translates leads to a weaker leaky attenuation constant, longer radiation aperture and hence more directive radiation. The radiation angle shift of 0.08° between the two beams in Fig. 9(b) is caused by the refractive index change in the silicon perturbations under carrier injection mode with excess carrier density of $10^{19}$ cm$^{-3}$, and refractive index change of $1.5\times10^{-2}$ correspondingly (Fig. 6c, top). Though weak, it has been demonstrated that the tuning effect can be further boosted by integrating the OLWA with resonator topologies [11], [20], [25], such as a Fabry-Pérot resonator or a ring-resonator. High-quality factor resonators can be put into or pushed out of resonance by slight variations in the guided wavelength achievable with carrier modulation in Si inclusions, which in turn leads to modulation of the OLWA's radiation with a large extinction ratio.

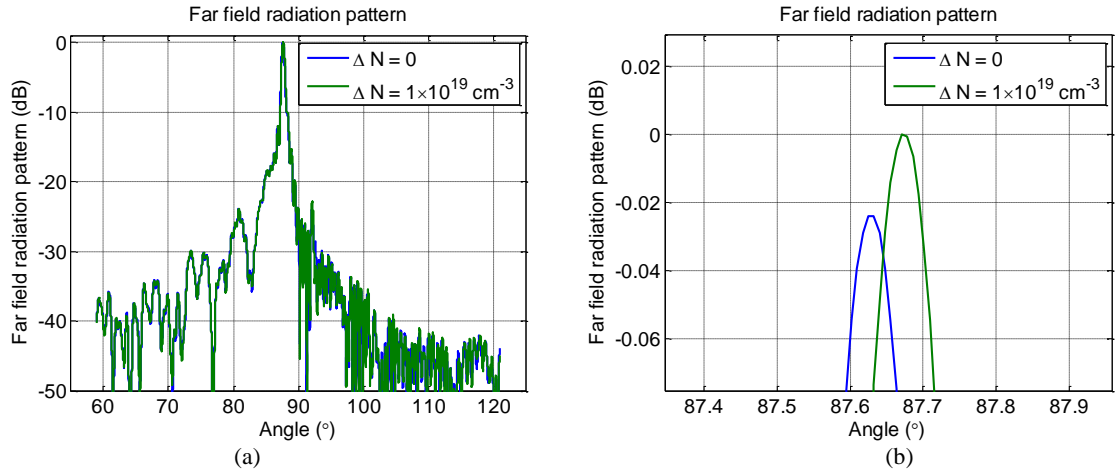

Figure 9. (a) Normalized far field radiation pattern in the *x-z* plane, assuming a two dimensional version of the OLWA as in [5], with carrier density change of $10^{19}$ cm$^{-3}$ in each silicon perturbation. (b) Radiation peak intensities and radiation peak locations assuming two cases with carrier density difference of $10^{19}$ cm$^{-3}$ in each silicon perturbation, in injection mode operation. The weak difference can be enhanced using the technique similar to that in [20] where the OLWA is placed inside a resonator.

## 5. CONCLUSION

Optical leaky-wave antennas (OLWAs) with fast tunable radiation pattern are attractive for integrated ultrafast optical devices. We fabricated an OLWA composed of $Si_3N_4$ waveguide on $SiO_2$ substrate and periodic subwavelength silicon perturbations. We also tested and characterized the radiation performance of the device. The measured far-field pattern is very directive and has the maximum intensity at the angle of 84.4° relative to the waveguide axis at the operative wavelength of 1540 nm. The optical antenna device has the potential for space-multiplexing integration, 3D photonics interconnect, spectrometer and planar imaging system. Moreover, the semiconductor based perturbations enables electrical tunability of the device. We assessed the electrically tunable range of refractive index change in Si domains of OLWA. By taking the response time of *p-n* junction width change *Δw,* and refractive index change *Δn* in the modulation region (Fig. 6b) as a figure of merit, the silicon wire modulator has a -3 dB bandwidth of 75 GHz with refractive index change of $3\times10^{-4}$ in carrier depletion mode, and 350 MHz bandwidth with refractive index change of $1.5\times10^{-2}$ in carrier injection mode.

## ACKNOWLEDGMENT

This work is supported by the National Science Foundation Award # ECCS-1028727. The authors would like to thank COMSOL, Inc. for providing COMSOL Multiphysics, and Lumerical Solutions, Inc. for providing DEVICE simulation tool.

*This manuscript has been submitted to 2015 SPIE Photonics West Proceedings.